\documentstyle[epsfig]{mn}

\begin{document}

\def\simlt{\mathrel{\rlap{\lower 3pt\hbox{$\sim$}}\raise 2.0pt\hbox{$<$}}}
\def\simgt{\mathrel{\rlap{\lower 3pt\hbox{$\sim$}} \raise 2.0pt\hbox{$>$}}}
\def\be{\begin{equation}}
\def\ee{\end{equation}}
\def\HI{\hbox{H$\scriptstyle\rm I\ $}}
\def\msun{M_{\odot}}
\def\cm3{\;\mbox{cm}^{-3}}
\def\erg{{\rm erg}}
\def\sec{{\rm s}}
\def\etal{{ et al.~}}
\def\ie{{\frenchspacing\it i.e. }}
\def\eg{{\frenchspacing\it e.g. }}

\title[Dust Formation in Very Massive Primordial Supernovae]
{Dust Formation in Very Massive Primordial Supernovae}
\author[Schneider, Ferrara \& Salvaterra]
{R. Schneider $^{1,2}$, A. Ferrara $^{3}$ \& R. Salvaterra $^{3}$  \\
$^1$INAF/Osservatorio Astrofisico di Arcetri, L.go E. Fermi 5, 50125 Firenze, Italy\\ 
$^2$Centro ''Enrico Fermi'', Via Panisperna 89/A, 00184 Roma, Italy \\
$^3$SISSA/International School for Advanced Studies, Via Beirut 4, 34100 Trieste, Italy}
 
\maketitle \vspace {7cm}
 
\begin{abstract}
At redshift $z\simgt 5$ Type II supernovae (SNII) are the only known dust sources with evolutionary 
timescales shorter than the Hubble time. We extend the model of dust formation in
the ejecta of SNII by Todini \& Ferrara (2001) to investigate the same process in 
pair-instability supernovae (PISN), which are though to arise from the explosion
of the first, metal free, very massive (140--260 $M_\odot$) cosmic stars. 

We find that 15\%-30\%
of the PISN progenitor mass is converted into dust, a value $>$~10 times higher 
than for SNII; PISN dust depletion factors (fraction of produced metals locked into dust grains) range between 
0.3 and 0.7. These conclusions depend very weakly on the mass of the PISN stellar progenitor,
which instead affects considerably the composition and size distribution. 
For the assumed temperature evolution, grain condensation
starts 150-200 days after the explosion; the dominant compounds for all progenitor masses are SiO$_2$ and 
Mg$_2$SiO$_4$ while the contribution of amorphous carbon and magnetite grains grows with progenitor mass; 
typical grain sizes range between  $10^{-3}$ and a 
few 0.1$\mu$m and are always smaller than 1$\mu$m.
We give a brief discussion of the implications of dust formation for the IMF evolution 
of the first stars, cosmic reionization and the intergalactic medium.

\end{abstract}

 
\section{Introduction}

In the recent years, dust has been recognized to have an increasingly important role in
our understanding of the near and distant Universe. The dramatic effect of dust at low and moderate 
redshifts has been immediately recognized when a recostruction of the cosmic star formation history
from rest-frame UV/visible emission was first attempted: dust grains absorb stellar light and re-emit it
in the FIR. Thus, even a tiny amount of dust extinction can lead to a severe underestimate of the 
actual star formation rate (Pettini \etal 1998; Steidel \etal 1999). New IR, FIR and submm facilities 
have revealed the existence of populations of sources, such as SCUBA $z \simgt 1$ sources, that are thought
to be dust-enshrouded star forming galaxies or AGNs (Smail \etal 1997; Hughes \etal 1998) or the Extremely
Red Objects, which are at least partly populated by dusty star-forming systems at $z \sim 1$ 
(Cimatti \etal 2002). Finally, dust plays a critical role in galaxy evolution, accelerating the formation
of molecular hydrogen (H$_2$), dominating the heating of gas through emission of photoelectrons 
in regions where UV fields are present and contributing to gas cooling through IR emission 
(see Draine 2003 for a recent thorough review). 

These observational evidences have motivated a series of studies aimed at including a treatment of
dust formation within galaxy evolution models (Granato \etal 2000; Hirashita \& Ferrara 2002; Morgan \&
Edmunds 2003). Since dust formation in the interstellar medium (ISM) is extremely inefficient (Tielens 1998), the preferential
sites of formation are considered to be the atmospheres of evolved low-mass ($M< 8 \msun$) stars, from where it is 
transported into the ISM through stellar winds (Whittet 1992). However, this mechanism can not explain 
the presence of dust at redshifts $> 5$ because at these high redshifts the evolutionary timescales of 
low-mass stars (0.1 -- 1 Gyr) start to be comparable with the age of the Universe.

Evidences for the presence of dust at high redshifts come from observations of damped Ly$\alpha$ systems  
(Pettini \etal 1994; Prochaska \& Wolfe 2002; Ledoux, Bergeron \& Petitjean 2002) and from the detection of dust thermal 
emission from high redshift QSOs selected from the SDSS survey out to redshift 5.5 and re-observed at mm wavelengths
(Omont \etal 2001; Carilli \etal 2001; Bertoldi \& Cox 2002). Very recently, Bertoldi \etal (2003) have reported the
observations of three $z>6$ SDSS QSOs at 1.2 mm,  detecting thermal dust radiation. From the IR luminosities, the estimated
dust masses are huge ($>10^8 \msun$) implying a high abundance of heavy elements and dust at redshifts as high as 
6.4 that can not be accounted by low-mass stars. Thus, dust enrichment must 
have occurred primarily on considerably shorter timescales in the ejecta of supernova 
explosions (Dwek \& Scalo 1980; Kozasa, Hasegawa \& Nomoto 1989; Todini \& Ferrara 2001; Nozawa \etal 2003).  

The strongest evidence for dust formation in supernova explosions was seen in SN 1987A (Moseley \etal 1989; Wooden 1997). 
In this event, the increased IR emission was accompanied by a a corresponding decrease in the optical emission and the emission-line
profiles were observed to shift toward the blue (McCray 1993 and references therein). 
In another supernova explosion, SN 1998S, the observed 
evolution of the hydrogen and helium line profiles argues in favor of dust formation within the ejecta as the redshifted 
side of the profile steadily faded while the blueshifted side remained constant (Gerardy \etal 2000). Dust emission has been seen 
in the supernova remnant Cassiopea A: similarly to SN 1987A, the total dust mass derived from IR luminosities is less than expected
from theory, suggesting that a colder population of dust grains may be present that emit at longer wavelengths and it is not detectable 
in the IR (Morgan \& Edmunds 2003; Dunne et al. 2003). Finally, in a considerable number of cases, supernovae have shown 
IR emission that was stronger toward longer wavelengths (see Gerardy \etal 2002 and references therein).
This IR excess has been generally interpreted as due to thermal emission from dust forming in 
the ejecta but alternative explanations exist; in particular, new IR observations of five Type II SNe have shown late-time emission 
that remains bright many years after the maximum and that it is hard to reconcile with emission from newly formed dust; 
IR echos from pre-existing dust in the circumstellar medium heated by the supernova flash might represent an alternative interpretation
(Gerardy \etal 2002).

Theoretical studies have started to investigate the process of dust formation in expanding SN ejecta. Most of the available models
are based on classical nucleation theory and grain growth (Kozasa \& Hasegawa 1987; Todini \& Ferrara 2001) with the exception of
the computations recently performed by Nozawa et al. 2003, who have 
made an extensive analysis of dust grain properties forming in zero-metallicity SNe of various masses. 
  The model developed by Todini \& Ferrara
(2001) is able to predict the dust mass and properties as a function of the initial stellar progenitor mass and metallicity. In spite
of the many uncertainties and approximations, this model has been shown to satisfactorily reproduce the observed properties of 
SN 1987A and of the young dwarf galaxy SBS 0335-052 (Hirashita, Hunt \& Ferrara 2002). 
 
In this paper, we investigate the formation of dust grains in the ejecta of very massive primordial supernovae, which are 
commonly known as pair-creation or pair instability SNe (PISN). These violent explosions are thought to terminate the life of 
metal-free stars with initial mass in the range $140 \msun \leq M \leq 260 \msun$ (Heger \& Woosley 2002). Indeed, detailed theoretical
modelling of the nucleosynthesis and internal structure of these very massive stars has shown that after central helium burning 
electron/positron pairs are created, converting a large fraction of internal energy into rest mass of the pairs; as a consequence,
the stars rapidly contract until explosive oxygen and silicon burning is able to revert the collapse and the stars are completely
disrupted in giant explosions (Fryer, Woosley \& Heger 2001). Below and above the mass range of PISN, the
most likely outcome of the evolution of metal-free stars appears to be black hole formation, cleanly separating the contribution
to metal and eventually dust production of such very massive stars from that of other mass ranges. 

We base our analysis on the model developed by Todini \& Ferrara (2001) and apply it to PISN. 
We therefore change the initial chemical composition of the ejecta and also the 
thermodynamical/dynamical properties which determine the ejecta evolution.
A larger amount of dust is expected to form in the ejecta of PISN with respect to Type II SNe,
because of the larger amount of metals released and the absence of
fallback of material onto the compact remnant (no remnant is expected to 
survive PISN explosions).

The motivation for the present study comes from the increasing number of evidences which seem to indicate that the first stars 
that were able to form in the early Universe from the collapse of metal-free gas clouds were indeed very massive, 
with characteristic masses of a few 100 $\msun$. These include detailed numerical simulations (Abel, Bryan \& Norman 2000, 2002; 
Bromm, Coppi \& Larson 1999, 2002; Bromm \etal 2001; Ripamonti \etal 2002), semi-analytic models 
(Omukai \& Nishi 1998; Nakamura \& Umemura 2001, 2002; Schneider \etal 2002) and several pieces of observations
(Hernandez \& Ferrara 2001; Oh \etal 2001; Salvaterra \& Ferrara 2002; Schneider \etal 2003). In particular, some of these studies have
pointed out that because of the reduced gas cooling efficiency, low-mass star formation is strongly inhibited before a minimum level 
of metal enrichment of the collapsing gas cloud has been reached (Bromm \etal 2001, Schneider \etal 2002). The value of this minimum
level, $Z_{\rm cr}$ is very uncertain, but likely to be between $10^{-6}$ and $10^{-4}   Z_{\odot}$ (Schneider \etal 2002; 2004).
Within this critical range of metallicities, the presence of dust appears to have a major role, providing an additional pathway for cooling
the gas, that fragments into lower mass clumps enabling the formation of second-generation low-mass stars (Schneider \etal 2003).
Thus, to estimate the efficiency of this dust-regulated low-mass star formation channel in very metal-poor gas clouds, it is crucial to
develop a model which is able to predict the amount of dust formed in the ejecta of PISN. 



\section{Dust Formation Model}

The process of dust formation and the resulting dust grain properties depend
on the physical conditions at the site of formation. In particular, several 
studies (see Kozasa, Hasegawa \& Nomoto 1989 and references therein) have 
shown that the time of onset of grain formation depends on the temperature 
structure in the supernova ejecta whereas the grain composition mainly 
reflects its chemical composition, which depends on the nucleosynthesis 
occurring during stellar lifetime (\ie on the progenitor mass) and explosion.

Thus, models of dust formation in supernova ejecta are based on specific 
prescriptions for the chemical composition and thermodynamics of the expanding 
gas. At the onset of shock generation, which occurs at the boundary of the 
innermost Fe-Ni core, the progenitor star is characterized by the standard 
stratified (onion-skin) structure. During shock propagation through the star, 
the gas undergoes a new phase of (explosive) nucleosynthesis and mixing of the 
internal layers is thought to occur at least up to the outer edge of the helium 
layer, as suggested by observations of early emergence of $\gamma$ and X-rays 
in SN 1987A (Kumagai \etal 1988 and references therein). These observations 
confirm that macroscopic mixing of the ejecta occurs through Rayleigh-Taylor 
instabilities but it is still very debated whether mixing is extended at molecular
level (Clayton et al. 1999; 2001).      
 
The model of Todini \& Ferrara (2001) (but see also Kozasa \etal 1989) is based 
on the assumption that materials are uniformly mixed from the center to the helium
outer edge. As a consequence, the temperature and density within this metal-rich
volume are assumed to be constant. When the shock reaches the surface of the 
progenitor, the star starts to expand and the expansion becomes homologous. Thus, 
the velocity of gas at a given layer is constant in time and proportional to the
radius from the center, \ie $v = R/t$, where $t$ is the time from the explosion and 
the expansion velocity is assumed to be
\be
v = \left(\frac{E_{\rm kin}}{M_{\rm tot}}\right)^{1/2}
\ee
where $E_{\rm kin}$ and $M_{\rm tot}$ are the kinetic energy and total mass ejected 
by the supernova. The temperature of the expanding ejecta is determined by various 
heating and cooling mechanisms. Todini \& Ferrara (2001) following 
Kozasa et al. (1989) assumed that when the gas reaches the photosphere, the gas
temperature is equal to the photospheric temperature and thereafter the temperature   
evolution follows from the
assumptions of adiabatic expansion for a perfect-gas, so that
\be
T = T_i \left(1 + \frac{v}{R_i} t \right)^{3 (1-\gamma)}.
\label{eq:T}
\ee
where $\gamma$ is the adiabatic index and the quantities $T_i$ and $R_i$ are the
photospheric temperature and radius obtained from the observational results of 
SN 1987A (Catchpole \etal 1987).
\begin{figure}
\center{{
\epsfig{figure=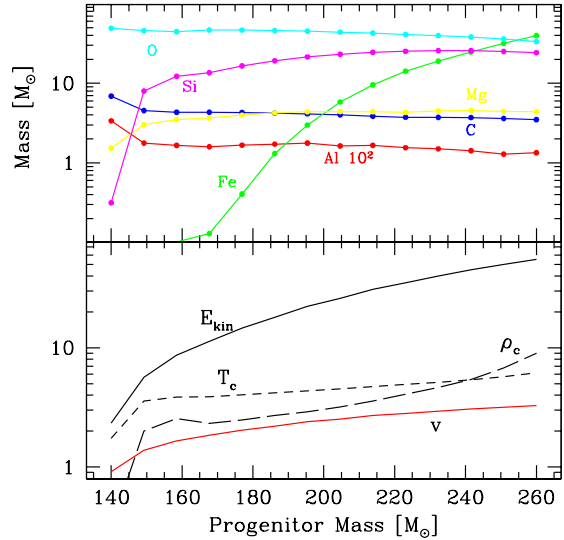,height=9cm }
}}
\caption{{\it Top panel}: dominant metal yields [$\msun$] of PISN
(Heger \& Woosley 2002) as a function of the progenitor stellar mass, 
after all unstable isotopes have decayed. 
{\it Bottom panel}: kinetic energy (in units $10^{51}$~erg), velocity of the
ejecta (in units $10^3$~km/s), central temperature (in units $10^9$~K) and 
density (in units $10^6$~gr/cm$^3$) as a function of the progenitor stellar
mass (see text).}
\label{fig:1}
\end{figure}
The lack of observational constraints for PISN forces us to model the thermal evolution
of the expanding ejecta on the basis of theoretical considerations. In particular, 
the evolution and internal structure of PISN have been studied
in great detail through numerical simulations (Fryer, Woosley \& Heger 2001; 
Heger \& Woosley 2002). Zero-metallicity progenitor stars with masses between 140 $\msun$
and $260 \msun$ evolve without significant mass loss until central helium burning. These
stars, after central helium depletion, have enough central entropy that they enter a 
temperature and density regime in which electron/positron pairs are created in abundance,
converting internal gas energy into rest mass of the pairs, without contributing much to the 
pressure. When this instability is encountered, the stars contract rapidly until explosive
oxygen and silicon burning produce enough energy to revert the collapse and the stars 
are completely disrupted in a giant explosion. The maximum central temperature and density during the
bounce as a function of the PISN progenitor mass 
together with the elemental composition of the emerging ejecta 
are shown in Fig.~\ref{fig:1}. The dominant metal yields show that nucleosynthesis in pair-creation supernovae
produce a total mass of O, Si, Mg and Al which is roughly independent of the initial progenitor mass but an Fe mass
which varies greatly with the mass of the progenitor, being almost negligible for initial stellar masses $< 200 \msun$.
It is important to stress that this should be interpreted as the iron mass only after all unstable isotopes have decayed. 
Indeed, unlike Type II SN progenitors, PISN progenitors do not build up an iron 
core before the explosion and the final Fe mass is generated through the 
decay of $^{56}$Co. We will consider this process in detail as this decay provides the relevant destruction process of CO
molecules.     

\begin{figure}
\center{{
\epsfig{figure=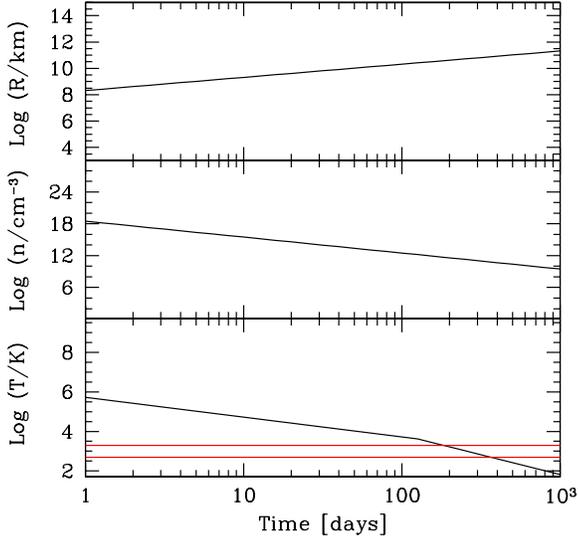,height=9cm }
}}
\caption{Dynamics and thermodynamics of the ejecta for a 195 $\msun$ $\,\,$ PISN.
The {\it top, medium} and {\it bottom} panels show the time evolution of the radius,  Log [R/km], 
number density, Log [n/cm$^3$], and temperature, Log [T/K], of the expanding gas. 
The temperatures corresponding to the onset of grain condensation ($T \leq 2000$~K) and to the
final state followed by the model ($T=500$~K) are indicated with two horizontal lines.}
\label{fig:2}
\end{figure}
The internal structure of the star at this maximum central density appears to have an approximately 
constant density, $\rho_c$, and temperature, $T_c$, up to the outer edge of the helium core (Fryer \etal 2001). 
We can adopt these values as the initial density and temperature of the volume where the metals are 
uniformly mixed at shock generation. Indeed, the time it takes the shocks to propagate through the 
low-density hydrogen envelope and reach the surface of the progenitor, starting the stellar expansion, 
is negligible. Adopting this 
normalization in Eq.~\ref{eq:T}, we can follow the evolution of the supernova ejecta. The results
are shown in Fig.~\ref{fig:2}. In the top and medium panels we show the evolution of radius and
number density of the ejecta of a 195 $\msun$ pair-creation supernova 
as a function of the time elapsed from the explosion. 
In the bottom panel, we show the corresponding evolution of the
temperature. Two different regimes can be identified: in the initial phase of the evolution, the 
ejecta is radiation-dominated and we assume an adiabatic index $\gamma=4/3$. This appears to be consistent
with the internal structure of the 250 $\msun$ star at maximum central density in the simulation of 
Fryer \etal (2001), which shows a density and temperature profile in the outer H-envelope compatible with
$\gamma \simeq 4/3$. However, for $t>111.26$~days (where 111.26 days is $^{56}$Co e-folding time) 
energy deposition from decaying radioactive elements is negligible and the gas cools adiabatically with
$\gamma=5/3$. A comparable thermal history is found by Nozawa \etal (2003) solving the radiative transfer
equation and taking into account the energy deposition by radioactive elements. For the same PISN progenitor
mass, however, their normalization is slightly higher (by a factor $\sim 3$) because of the different
input stellar models (Umeda \& Nomoto 2002).  

\section{Model Results}

In the model of Todini \& Ferrara (2001), the formation of dust grains from a gaseous 
metal-rich medium is described as a two step process: 
(i) the formation of ``critical clusters'' at the corresponding condensation barrier and 
(ii)the subsequent growth of these clusters into macroscopic dust grains through mass accretion.
 A thorough description of the model, which follows the classic theory of nucleation, can be found in the paper 
of Todini \& Ferrara (2001) to which we refer interested readers. 

In the present study, we follow the formation of seven solid compounds, namely, Al$_2$O$_3$ (corundum), iron, 
Fe$_3$O$_4$ (magnetite), MgSiO$_3$ (enstatite), Mg$_2$SiO$_4$ (forsterite), SiO$_2$ and amorphous carbon 
(AC) grains. The corresponding chemical reactions are listed in Table~1. Numerical constants can be found in
Table~1 of Todini \& Ferrara (2001) except for SiO$_2$, for whom we have adopted the values used by Nozawa \etal (2003). 

\begin{table*}
\caption{\footnotesize Chemical reactions included in dust formation calculations.}
\begin{center}
\begin{tabular}{@{}|l|c|@{}} \hline
Solid compound &        Chemical reaction             \\ \hline       
Al$_2$O$_3$    &  2Al+3O $\rightarrow$ Al$_2$O$_3$        \\
Fe             &  Fe(g)  $\rightarrow$ Fe(s)              \\
Fe$_3$O$_4$    &  3Fe+4O $\rightarrow$ Fe$_3$O$_4$        \\
MgSiO$_3$      &  Mg+SiO+2O $\rightarrow$ MgSiO$_3$       \\
Mg$_2$SiO$_4$  &  2Mg+SiO+3O $\rightarrow$ Mg$_2$SiO$_4$  \\
SiO$_2$        &  SiO+O $\rightarrow$ SiO$_2$             \\
AC             &  C(g) $\rightarrow$ C(s)                 \\ \hline
\end{tabular}
\end{center}
\end{table*}

It is important to note that, as we will see, AC grains can form also
if the ejecta composition is richer in oxygen than in carbon (O $>$ C) as
Clayton, Liu \& Dalgarno (1999) have shown.
The formation of CO and SiO molecules in SN ejecta can be very important for dust formation, 
because carbon atoms bound in CO molecules are not available to form AC grains and SiO molecules take 
part in the reactions which lead to the formation of MgSiO$_3$, Mg$_2$SiO$_4$ and SiO$_2$ (see Table~1). 
Thus, the process of molecule formation in the expanding ejecta is followed at temperatures $T \leq 2 \times 10^4$~K together with
the $^{56}{\rm Co} \rightarrow$ $^{56}{\rm Fe}$ radioactive decay as the impact with energetic electrons produced during 
this decay represents the main destruction process of CO molecules.    

\begin{figure}
\center{{
\epsfig{figure=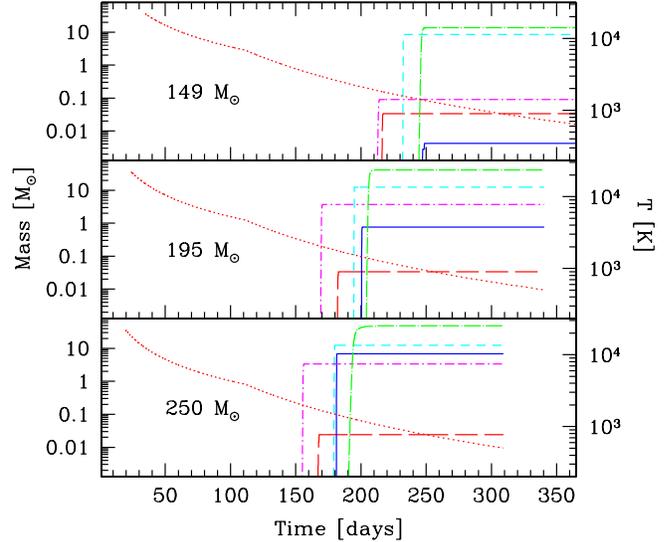,height=9cm }
}}
\caption{Time evolution of the dust mass in different solid compounds synthetized by
a $149 \msun$ ({\it top panel}), $195 \msun$ ({\it medium panel}) and $250 \msun$ ({\it bottom panel}) 
PISN. The different lines correspond to  AC (dot-dashed), Al$_2$O$_3$ (long-dashed), Mg$_2$SiO$_4$ (dashed),
Fe$_3$O$_4$ (solid)  and SiO$_2$ (dot-long dashed). 
The dotted lines indicate the corresponding values of the 
temperature of the ejecta.}
\label{fig:3}
\end{figure}

The formation of dust grains is followed until the temperature of the ejecta 
has decreased to about $500$~K, because at this stage condensation processes are terminated for all solid compounds.

The time evolution of the dust mass synthetized in different compounds is shown in Fig.~\ref{fig:3} 
for three representative PISN progenitor masses, $149 \msun$, $195 \msun$ and $250 \msun$. Dust grains 
start to condensate when the temperature of the expanding gas has dropped below 2000~K, about 150-200 days after the explosion 
(see dotted lines).
During this time interval, a fraction of the $^{56}{\rm Co}$ initially present in the ejecta has decayed to Fe. Thus, at the onset of grain condensation, some C atoms are bound in CO molecules and
others are available to form amorphous carbon grains. The resulting dust mass is mainly composed by silicates and magnetite grains.  
Amorphous carbon grains are the first to form, followed by corundum, forsterite, magnetite and ultimately SiO$_2$.
This evolutionary sequence simply reflects the different condensation temperatures. 
The relative abundances of the different grain species depend on the progenitor mass: the $149 \msun$ star ejects less iron 
($^{56}{\rm Co}$) than higher mass PISN and thus its final dust composition is characterized by a lower concentration of 
magnetite and carbon grains (the latter because most of the initial carbon grains are locked in CO molecules).

\subsection{Final dust masses}

\begin{figure}
\center{{
\epsfig{figure=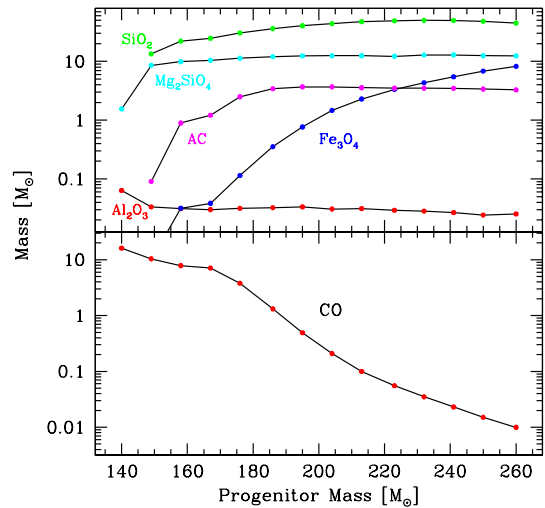,height=9cm }
}}
\caption{{\it Top panel:} final dust mass of different solid compounds formed in PISN ejecta 
as a function of the mass of the stellar progenitors.
{\it Bottom panel:} total mass of CO molecules synthetized in the ejecta as a function of the initial mass 
of the stellar progenitors.}
\label{fig:4}
\end{figure}
The total dust mass formed in different solid compounds and the total mass of CO molecules synthetized in the ejecta 
as a function of the progenitor stellar mass are shown in Fig.~\ref{fig:4}. For all but the smallest progenitor star,
silicate grains appear to be the dominant dust compounds. Indeed, all SiO molecules initially present in the ejecta 
take part in the reactions leading to the formation of forsterite and SiO$_2$ grains and only CO molecules are left
at the end of dust condensation. 
At the lowest mass bin, $140 \msun$, the initial $^{56}{\rm Co}$ mass is not large enough to favour the formation 
of magnetite and amorphous carbon grains and only Al$_2$O$_3$ and Mg$_2$SiO$_4$ grains are formed. 
At higher initial stellar masses, the larger $^{56}{\rm Co}$ mass ejected by the PISN favours the formation of carbon and
magnetite grains and the final mass of CO molecules is correspondingly reduced. Note that MgSiO$_3$ grains are never formed
because of the highest condensation temperature and largest nucleation current of Mg$_2$SiO$_4$ grains, which lock all the 
Mg initially present in the ejecta. Similarly, Fe(s) are never formed because all the Fe which is produced by radioactive 
decays is locked into magnetite grains.       
Note that we do not consider the formation of MgO grains which, however, can occur only in the ejecta of the 
$140 \msun$ PISN where Si is less abundant than Mg, and only 40\% of the initial Mg mass is depleted onto 
Mg$_2$SiO$_4$ grains.

\subsection{Grain size distribution}

\begin{figure}
\center{{
\epsfig{figure=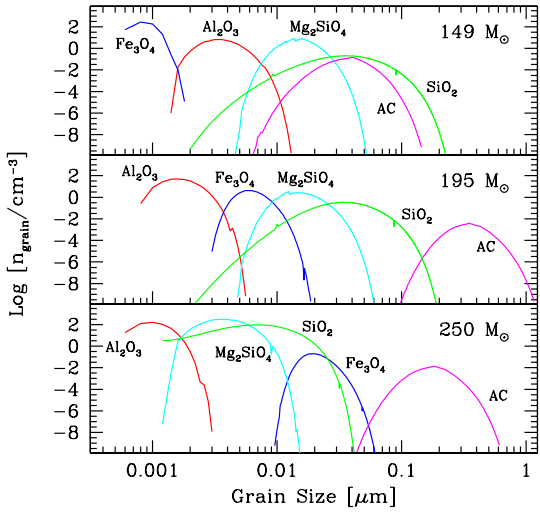,height=9cm }
}}
\caption{Final grain size distributions for different solid compounds synthetized by
a $149 \msun$ ({\it top panel}), $195 \msun$ ({\it medium panel}) and $250 \msun$ ({\it bottom panel}) 
PISN. The distribution is computed when the temperature of the ejecta has decreased to 500 K and formation and
accretion are terminated for all compounds.}
\label{fig:5}
\end{figure}

The final grain size distributions are shown in Fig.~5 for 149 $\msun$, 195 $\msun$ and 250 $\msun$ PISN. 
The characteristic 
grain sizes range between $10^{-3}$ and a few 0.1 $\mu$m, depending on the grain species and on the mass 
of the progenitor star.
Each dust grain grows by accretion to a final size which depends on the density regime when 
condensation occurs (higher densities favoring higher accretion rates) and on the abundance of the key species.  
Indeed, being the first to condense, amorphous carbon grains tend to have characteristic sizes larger than silicates
and magnetite grains. Al$_2$O$_3$ grains are typically small because of the rather low-abundance of Al in the 
ejecta. Larger mass progenitor systems tend to have larger magnetite grains and smaller amorphous carbon grains. However,
the properties of grain size distributions for different progenitor masses depend on the different thermodynamics of 
the expanding ejecta which affects the relative amplitude of cooling and nucleation rates.
The model predicts that, for all progenitor masses, PISN lead to the formation of grains with radii smaller than 1 $\mu$m
but systematically larger than the typical sizes of grains formed in SNII explosions (Todini \& Ferrara 2001) 
with a reduced scatter among different grain species.

\subsection{Depletion factors}

\begin{figure}
\center{{
\epsfig{figure=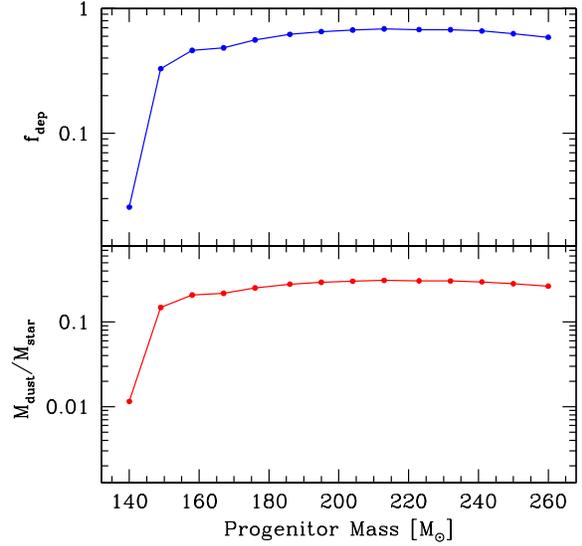,height=9cm }
}}
\caption{{\it Top panel}: total dust depletion factor defined as the dust-to-metal mass ratio as a function of the initial
progenitor mass. 
{\it Bottom panel}: corresponding values of the ratio between the total dust mass synthetized and the initial progenitor mass.}
\label{fig:6}
\end{figure}

The cosmological relevance of dust synthetized in PISN explosions will depend mainly on the global 
properties of dust rather than on the nature and size of different compounds. 
In Fig.~\ref{fig:6} we show the total dust depletion factor, defined as the dust-to-metal mass ratio
released in the explosions, $f_{\rm dep}=M_{\rm dust}/M_{\rm met}$ as a function of PISN progenitor mass. 
With the exception of the smallest mass bin, $f_{\rm dep}$ ranges between 0.3 and 0.7 and tends to increase with 
increasing progenitor mass. In the bottom panel of the same figure, we show the ratio of the total dust mass
and the initial progenitor stellar mass. For all but the smallest mass bin, the amount of dust formed is between
15 \% and 30 \% of the initial stellar mass. The peculiar behaviour of the 140 $\msun$ progenitor is due to the
original elemental composition of its ejecta, which is highly underabundant in Fe and Si and therefore only 
corundum is synthetized in the final ejecta (see also Fig.~4).

\begin{figure}
\center{{
\epsfig{figure=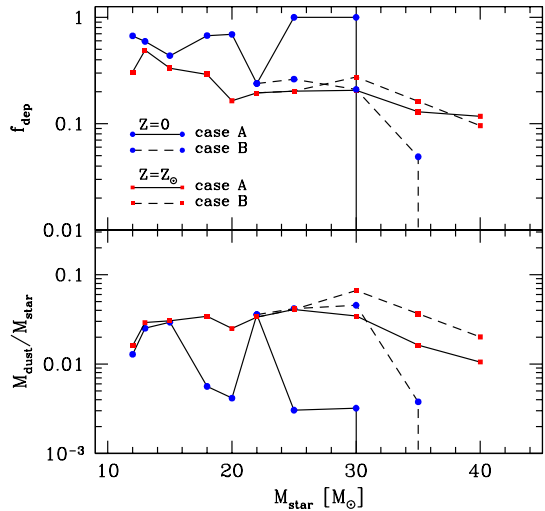,height=9cm }
}}
\caption{Same as in Fig.~\ref{fig:7} but for Type II SNe. Case A and B refer to low ($1.2 \times 10^{51} \erg$) and high 
($1.9 \times 10^{51} \erg$) kinetic energy scenarios and the dots (squares) represent the dust mass formed from 
Type II SNe with initial zero (solar) metallicity.}
\label{fig:7}
\end{figure}

For comparison, in Fig.~\ref{fig:7} we show the same quantities but for SNII. 
Following Todini \& Ferrara (2001), we show two cases corresponding to two different values of the total 
kinetic energy released in the explosion: case A corresponds to the low kinetic energy model 
($1.2 \times 10^{51} \erg$) and case B refers to the high kinetic energy model 
($1.9 \times 10^{51} \erg$) (see also the original paper of Woosley \& Weaver 1995). 
We also assume two different values for the
initial metallicity of the progenitor stars, $Z=0$ (dots) and $Z=Z_{\odot}$ (squares).

The moderate metal yields in SNII and the effect of fallback of material after 
the explosion onto the compact remnant lead to depletion factors 
which are comparable to PISN, with values ranging from 20 \% up to 70 \% 
for $Z=0$ progenitors with  masses $< 22 \msun$. For larger masses, 
the depletion factor decreases in case B scenarios (higher kinetic energy) 
because of the larger amount of metals released. For case A scenarios (lower kinetic energy),
instead, 25 $\msun$ and $30 \msun$ SNII are predicted to have an $f_{\rm dep}=1$. 
This is due to the fact that because of fallback, these stars eject only a rather small 
amount of metals, mostly in the form of carbon, which is completely 
depleted into AC grains. If the stellar progenitors have solar metallicities, 
the depletion factor ranges between $10\%$ and $50\%$
with a much reduced scatter between case A and B.

In spite of these large depletion factors, the total mass of dust synthetized by PISN 
is significantly higher than that produced by SNII.
Indeed, as already discussed in Todini \& Ferrara (2001), in case A scenarios, 
$Z=0$ SNII synthetize a total dust mass which corresponds to 
a fraction between 0.3 and 3\% of the original stellar progenitor mass (thus, $\sim 0.1 - 0.6 \msun$ of dust per SN).  
These values are slightly larger if case B models are considered. 
When the initial stellar progenitors have solar metallicity, the resulting dust mass is typically a factor 
$\sim 3$ larger than for the metal-free case but it is always less than 6 \% of the initial stellar mass.

\section{Summary and Discussion}

We have investigated the process of dust formation in the ejecta of first stellar explosions, assuming that 
the first stars form with characteristic masses of a few 100 $\msun$ and explode as pair-creation SNe. 
The study is based on an extension of the model developed by Todini \& Ferrara (2001) for Type II SNe, which
accounts for the different initial chemical compositions of the ejecta and  dynamical/thermodynamical
properties of the ejecta evolution. 

The main results of our analysis can be summarized as follows:

\begin{enumerate}
\item During pair-creation SN explosions, a significant amount of dust is synthetized out of the heavy elements
produced in the progenitor stellar interiors during the main sequence lifetime and at the onset of the explosion
due to explosive nucleosynthetic processes. The fraction of the original stellar mass which is converted into
dust depends on the progenitor mass, with typical values ranging between 
15\% and 30\%, resulting in $30 \msun$ - $60 \msun$ of dust produced per SN. 
These values are much larger than those found for $Z=0$ Type II SNe resulting from stars with masses between 
12 $\msun$ and 40 $\msun$ ($0.1 \msun$ - $0.6 \msun$), even if these stars are assumed to be of solar metallicity 
($0.1 \msun$ - $1.8 \msun$).  

\item Because of the large amount of metals released in PISN explosions, dust depletion factors, defined as the 
dust-to-metal mass ratios, range between  30\% and 70\% depending on  the progenitor mass.

\item The composition of the dust compounds depends critically on the thermodynamics and initial composition 
of the ejecta, \ie on the progenitor mass. 
For the assumed temperature evolution, grain condensation starts 150-200 days after the explosion and 
silicates, magnetite and AC grains are formed. The dominant compounds for all progenitor masses are SiO$_2$ and 
Mg$_2$SiO$_4$ while the contribution of amorphous carbon and magnetite grains grows with progenitor mass. 

\item PISN explosions lead to the formation of grains with typical sizes which range between $10^{-3}$ and a 
few 0.1$\mu$m; for all progenitors, grains radii are always smaller than 1 $\mu$m but the relative grain size 
distribution change depending on the thermodynamics and abundance of the relevant key species. As a general trend,
corundum represents the smallest-size compound, amorphous carbon grains tend to form with the largest radii and 
silicates and magnetite grains have intermediate sizes. 
 
\end{enumerate}

The above results have been obtained under the assumption that the heavy elements present in the ejecta at the 
onset of the explosion are uniformly mixed up to the outer edge of the helium core. 
In previous dust formation models (Kozasa \etal 1989; Todini \& Ferrara 2001) this approximation was motivated 
by the early emergence of X-rays and $\gamma$-rays in SN 1987A, which suggested mixing in the ejecta at least 
up to the helium core edge (Kumagai \etal 1988). The use of multi-dimensional hydrodynamic codes to model 
the observed light curves has made clear that mixing occurs on macroscopic scales through the development 
of Rayleigh-Taylor instabilities: these instabilities arise at the interface between elemental
zones and grow non-linearly to produce (i) fingers of heavy elements projected outwards with high velocities and (ii) 
mixing of lighter elements down to regions that have lower velocities (Wooden 1997). 
As a result, the gas in the ejecta is mixed into regions which are still chemically homogeneous and which cool with 
different timescales, whereas only small clumps in the ejecta are microscopically mixed.
Such a structure affects the process of dust formation, changing both the total amount of dust formed and the relative 
abundance of different solid compounds, as recently shown by Nozawa et al. (2004).

A related aspect of the model which requires deeper investigation is the assumed temperature structure of the gas 
within the ejecta. The adopted thermal evolution of the expanding ejecta is motivated by the results of numerical
simulations of the internal structure of PISN progenitors (Fryer, Woosley \& Heger 2001; Heger \& Woosley 2002) 
and models the impact of the decays of radioactive elements through the change of the adiabatic index from 
$\gamma = 4/3$ to $\gamma=5/3$ assumed to occurr at  $^{56}$Co e-folding time (111.26 days after the explosion).
A comparable thermal history is found by Nozawa \etal (2003) solving the radiative transfer
equation and taking into account the energy deposition by radioactive elements. For the same PISN progenitor
mass, however, their normalization is slightly higher (by a factor $\sim 3$) because of the different
input stellar models (Umeda \& Nomoto 2002).

Finally, to quantify the cosmic relevance of dust formation in the early Universe, we should restrict the 
analysis to the fraction of the newly formed dust which is able to survive the impact of the reverse shock, 
following thereafter the fate of the surrounding metal-enriched gas. The process of dust sputtering dissociates 
the grains into their metal components and might have an important role in cosmic metal enrichment 
(Bianchi \& Ferrara 2004). However, new theoretical models seem to indicate that the post-shock temperature 
enters the regime suitable for dust condensation inside the oxygen layer, because of the high cooling efficiency 
of this element, but remains substantially higher in the outer He and H layers. If this is the case, then the 
reverse shock might lead to an increase in the total amount of dust formed rather than decreasing it.

\section{Cosmological Implications}

Our analysis shows that if the first stars formed according to a top-heavy IMF and a fraction of them exploded 
as pair-creation supernovae, a large amount of dust is produced in the early Universe. In this section, we discuss 
some of the main cosmological implications of this result, with particular emphasis on its role in the 
thermodynamics of the gas that will be later incorporated into subsequent generations of objects.

Recent numerical and semi-analytical models for the collapse of star-forming gas clouds in the early Universe 
have shown that because of the absence of metals and the reduced cooling ability of the gas, the formation of 
low-mass stars is strongly inhibited.
In particular, below a critical threshold level of metallicity of $Z_{\rm cr} = 10^{-5 \pm 1} Z_{\odot}$ cooling and 
fragmentation of the gas clouds stop when the temperature reaches a few hundreds K (minimum temperature for 
H$_2$ cooling) and the corresponding Jeans mass is of the order of $10^3 - 10^4 \msun$ (Schneider \etal 2002). 
Gas clouds with mass comparable to the Jeans mass start to gravitationally collapse without further fragmentation, 
until a central protostellar core is formed which rapidly grows in mass through gas accretion from the surrounding 
envelope. The absence of metals and dust in the accretion flow and the high gas temperature favour very high 
accretion efficiencies and the resulting stars can be as massive as 600 $\msun$ (Omukai \& Palla 2003).

As the gas becomes more and more enriched with heavy elements, the cooling rate increases because of metal 
(especially C and O) line emission. More importantly, if a fraction of the available metals is depleted onto 
dust grains, dust-gas thermal exchages activate a new phase of cooling and fragmentation which enables the 
formation of gas clumps with low-mass (Schneider \etal 2003). 
In particular, if the metallicity of the star forming gas clouds exceeds $10^{-4} Z_{\odot}$, this dust-driven 
cooling pathway is irrelevant because cooling via metal-line emission by itself is able to fragment the gas 
down to characteristic Jeans masses in the range $10^{-2}-1 \msun$. At the same time, if the metallicity of 
the gas is below $10^{-6} Z_{\odot}$, even if all the available metals are assumed to be depleted onto dust 
grains, the resulting cooling efficiency is too low to activate fragmentation and the resulting Jeans 
masses are as large as in the metal-free case ($10^3 - 10^4 \msun$). Thus, we can conclude that the presence 
of dust is crucial to determine the final mass of stars forming out of gas clouds with metallicities in the 
critical range $Z_{\rm cr} = 10^{-5 \pm 1} Z_{\odot}$. As a reference value, for a metallicity of 
$Z=10^{-5.1} Z_{\odot}$, if 20\% of the metals are depleted onto dust grains ($f_{\rm dep}=0.2$) the 
resulting mass of the collapsing clouds is reduced to $10^{-2}-10^{-1} \msun$
and can lead to low-mass stars. This is fully consistent with the predicted dust mass formed from 
PISN progenitors (see Fig.~6).

Also, we can not neglect that our models predict the formation of a significant amount of CO molecules, 
which can contribute to gas cooling. This aspect needs to be investigated further though we expect that the 
presence of CO molecules will be complementary to that of dust in the critical range of metallicities.
 
Therefore, in the emerging picture of galaxy evolution, the first episodes of star formation were characterized 
by very massive stars forming according to a top-heavy IMF. It is only when the metals and dust ejected by the first 
pair-creation supernovae are able to pollute a substantial fraction of the IGM, that an overall transition to a normal 
IMF forming stars with masses comparable to those that we presently observe in the nearby Universe occurs. 
The epoch of this transition depends crucially on the process of metal enrichment through the so-called 
chemical feedback: within metal-enriched regions, 
if the metallicity lies within the critical range, the presence of dust becomes critical 
and can no longer be neglected in the thermodynamics of the star forming gas. It is very likely that the transition 
will not occur at a single redshift because of the highly inhomogeneous process of metal enrichment 
(Scannapieco, Schneider \& Ferrara 2003) and that there will be epochs when the two modes of star formation will be 
coeval in different regions of the Universe.

This in turns might be very important for the reionization history of the IGM. 
Indeed, very massive metal-free stars are powerful sources of ionizing photons and an 
early epoch of star formation with a top-heavy IMF can easily match the required high optical depth
to electron scattering measured by WMAP (Kogut \etal 2003; Spergel \etal 2003; 
for a general discussion see Ciardi, Ferrara \& White 2003). 
However, an early epoch of reionization is difficult to reconcile with the observed Gunn-Peterson 
effect in the spectra of $z>6$ quasars (Fan \etal 2002) which implies a mass averaged neutral 
fraction of $\sim 1$\%. These two observational constraints seem to indicate that the reionization
history of the Universe might have been more complex than previously thought, with probably two distinct 
epochs of reionization separated by a relatively small redshift interval in which the Universe might have 
recombined again. This intermediate epoch of recombination could be the result of a decrease in the 
ionizing power of luminous sources, probably due to their metal-enrichment and consequent IMF 
transition. 

\begin{figure}
\center{{ 
\epsfig{figure=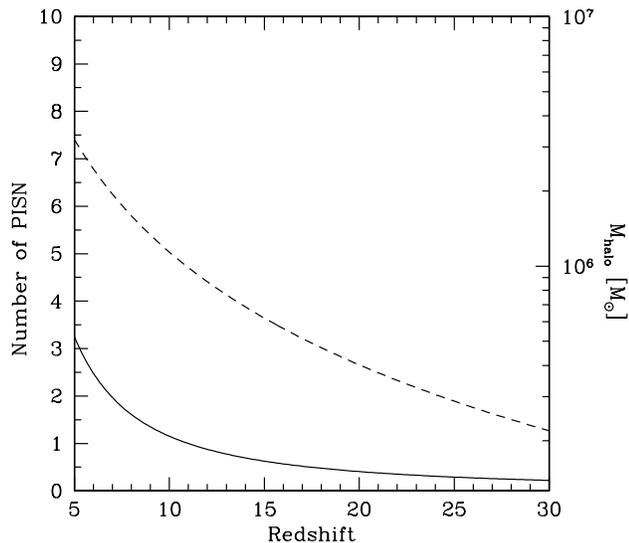,height=9cm }
}}
\caption{The number of PISN required to pollute a host protogalaxy with a dust-to-gas ratio equal to 5 \% of the galactic value
is shown as a function redshift (solid line). In the same panel, we show the total mass of the host halo which correspond to a 
virial temperature of 1000~K (dashed line).}
\end{figure}
An early epoch of reionization, as required by WMAP data, poses another critical issue: after reionization, 
the temperature of the IGM starts to decline as a consequence of cosmic expansion. By the time it reaches the 
observable range of redshifts $z<6$ the temperature of the IGM might be too low to match the observed thermal 
history (Schaye \etal 2000). The presence of a significant amount of dust synthetized by the first very massive 
supernovae might be extremely important to raise the IGM temperature through dust photoelectric heating. 
To estimate the amount of dust in the IGM required to match the observed thermal history, it is necessary to 
make specific assumptions about the UV background radiation, the reionization history of the IGM
and the grain composition and size distribution (Inoue \& Kamaya 2003). Furthermore, the properties of dust grains in the
IGM may differ from those directly predicted by dust formation models as a consequence of specific selection rules 
in the transfer of grains from the host galaxies to the IGM (see Bianchi \& Ferrara 2004, who find that only grains
larger than $\approx 0.06 \mu$m are preferentially ejected in the IGM). 
These complications are beyond the scope of the present analysis.

Finally, it is well known that the presence of dust at high redshifts offers an alternative formation 
channel for molecular hydrogen, the dominant coolant in the early Universe, which, in the absence of dust, 
can form only from the gas phase. In small protogalaxies, the H$_2$ formation rate on grain surface becomes 
dominant with respect to the formation rate from the gas phase, when the dust-to-gas ratio exceeds roughly 
5\% of the galactic value (Todini \& Ferrara 2001). 
From this order of magnitude estimate, we can derive the number of pair-creation supernovae required to 
enrich a small protogalaxy to this level. The results are shown in Fig.~8, where we have assumed the host 
galaxy to correspond to a halo with virial temperature 
of 1000~K and we have considered the dust mass synthetized by a 195 $\msun$ progenitor 
(see Fig.~4) as representative of dust formation efficiencies in PISN. This simple estimate shows that in the 
redshift range relevant to these small protogalactic systems ($z>10-15$), one PISN is required to pollute 
the gas with enough dust that H$_2$ formation on grain surface starts to be important. 
This, in turns, might have very important consequences for the star formation activity at high redshift 
(Hirashita \& Ferrara 2002).

\section*{Acknowledgements}
This work was based on the code developed by P. Todini, whose support we wish to gratefully acknowledge. We also thank 
A. Heger and T. Kozasa for fruitful information and the anonymous referee for valuable suggestions and careful 
reading of the paper. We acknowledge support from the CNR/JSPS Italy-Japan Seminar Program.


\begin{thebibliography}{10}
\bibitem[]{}Abel T., Bryan G. L., Norman M. L. 2000, ApJ, 540, 39
\bibitem[]{}Abel T., Bryan G. L., Norman M. L. 2002, Science, 295, 93
\bibitem[]{}Bertoldi, F. \& Cox, P. 2002, A\&A, 884, L11
\bibitem[]{}Bertoldi, F.  \etal 2003, A\&A, 406, L55
\bibitem[]{}Bianchi, S. \& Ferrara, A. 2004, in preparation
\bibitem[]{}Bromm V., Coppi P. S., Larson R. B. 1999, ApJ, 527, L5
\bibitem[]{}Bromm V., Ferrara A., Coppi P. S., Larson R. B. 2001, MNRAS, 328, 969
\bibitem[]{}Bromm V., Coppi P. S., Larson R. B. 2002, ApJ, 564, 23
\bibitem[]{}Carilli, C., Bertoldi, F., Rupen, M. \etal 2001, ApJ, 555, 625 
\bibitem[]{}Catchpole, R. M. \etal 1987, MNRAS, 229, 15
\bibitem[]{}Ciardi, B., Ferrara, A. \& White, S. D. M. 2003, MNRAS, 344, L7
\bibitem[]{}Clayton, D. D., Liu, W. \& Dalgarno, A. 1999, Science, 283, 1290
\bibitem[]{}Clayton, D. D., Deneault, E. A. N., Meyer, B., S. 2001, ApJ, 562, 480
\bibitem[]{}Cimatti, A., Daddi, E., Mignoli, M. \etal 2002, A \& A, 381, L68
\bibitem[]{}Draine, B. T. 2003, to appear in ARAA vol 41 
\bibitem[]{}Dunne, L., Eales, S., Ivison, R., Morgan, H., Edmunds, M. 2003, Nature, 424, 285
\bibitem[]{}Dwek, E. \& Scalo, J. M. 1980, ApJ, 239, 193
\bibitem[]{}Fryer, C. L., Woosley, S. E. \& Heger, A. 2001, ApJ, 550, 372
\bibitem[]{}Gerardy, C. L., Fesen, R. A., H{\"o}flich, P. \& Wheeler, J. C. 2000, AJ, 119, 2968
\bibitem[]{}Gerardy, C. L. \etal 2002, ApJ, 575, 1007
\bibitem[]{}Granato, G. L., Lacey, C. G., Silva, L., Bressan, A., Baugh, C. M., Cole, S., Frenk, C.S. 2000, ApJ, 542, 710
\bibitem[]{}Heger A. \& Woosley S. E. 2002, ApJ, 567, 532
\bibitem[]{}Hernandez, X. \& Ferrara, A. 2001, MNRAS, 324, 484
\bibitem[]{}Hirashita, H. \& Ferrara, A. 2002, MNRAS, 337, 921
\bibitem[]{}Hirashita, H., Hunt, L. K. \& Ferrara, A. 2002, MNRAS, 330, L19
\bibitem[]{}Hughes, D. M. \etal 1998, Nature, 394, 241  
\bibitem[]{}Inoue, A. K. \& Kamaya, H. 2003, MNRAS, 341, L7
\bibitem[]{}Kogut, A.  \etal 2003, ApJS, 148, 175
\bibitem[]{}Kozasa, T. \&  Hasegawa, H. 1987, Prog. Theor. Phys., 77, 1402
\bibitem[]{}Kozasa, T., Hasegawa, H., Nomoto, K. 1989, ApJ, 344, 325
\bibitem[]{}Kumagai, S., Itoh, M., Shigeyama, T., Nomoto, K., Nishimura, J. 1988, A \& A, 197, L7
\bibitem[]{}Ledoux, C., Bergeron, J., Petitjean, P. 2002, A\&A, 385, 802 
\bibitem[]{}McCray, R. 1993, ARA\&A, 31, 175
\bibitem[]{}Morgan, H. L. \& Edmunds, M. G. 2003, MNRAS, 343, 427
\bibitem[]{}Moseley, S. H., Dwek, E.,Giaccum, W., Graham, J. R., Loewenstein, R. F., Silverberg, R. F. 1989, Nature, 340, 697
\bibitem[]{}Nakamura F. \& Umemura M. 2001, ApJ, 548, 19
\bibitem[]{}Nakamura F. \& Umemura M. 2002, ApJ, 569, 549
\bibitem[]{}Nozawa, T.,  Kozasa, T., Umeda, H., Maeda, K., \& Nomoto, K. 2003, ApJ, 598, 785
\bibitem[]{}Oh, S. P., Nollett, K. M., Madau, P. \& Wasserburg, G. J. 2001, ApJ, 562, L1
\bibitem[]{}Omont, A., Cox, P., Bertoldi, F., McMahon,R. G., Carilli, C., Isaak, K. G. 2001, A\&A, 374, 371
\bibitem[]{}Omukai, K. \& Nishi, R. 1998, ApJ, 508, 141
\bibitem[]{}Omukai K. \& Palla F. 2003,  ApJ, 589, 677
\bibitem[]{}Pettini, M., Smith, L. J., Hunstead, R. W. \& King, D. L. 1994, ApJ, 426, 79
\bibitem[]{}Pettini, M., Kellogg, M., Steidel, C., Dickinson, M., Adelberger, K. L., Giavalisco, M. 1998, ApJ, 508, 539
\bibitem[]{}Prochaska, J. X. \& Wolfe, A. M. 2002, ApJ, 566, 68
\bibitem[]{}Ripamonti, E., Haardt, F., Ferrara, A. \& Colpi, M. 2002, MNRAS, 334, 401
\bibitem[]{}Salvaterra, R. \& Ferrara, A. 2003, MNRAS, 339, 973
\bibitem[]{}Scannapieco, E., Schneider, R. \& Ferrara, A. 2003, ApJ, 589, 35
\bibitem[]{}Schaye, J., Theuns, T., Rauch, M., Efstathiou, G., Sargent, W. L. W. 2000, MNRAS, 318, 817
\bibitem[]{}Schneider, R., Ferrara, A., Natarajan, P., Omukai, K. 2002, ApJ, 571, 30
\bibitem[]{}Schneider, R., Ferrara, A., Salvaterra, R., Omukai, K., Bromm, V. 2003a, Nature, 422, 869
\bibitem[]{}Schneider, R., Ferrara, A., Salvaterra, R. \& Omukai, K. 2004, MNRAS, submitted
\bibitem[]{}Smail, I., Ivison, R. J., Blain, A. W. 1997, ApJ, 490, L5
\bibitem[]{}Spergel, D. L. et al. 2003, ApJS, 148, 175 
\bibitem[]{}Steidel, C. C., Adelberger, K.L., Giavalisco, M., Dickinson, M., Pettini, M. 1999, ApJ, 519, 1
\bibitem[]{}Tielens, A. G. G. M. 1998, ApJ, 499, 267
\bibitem[]{}Todini, P. \& Ferrara, A. 2001, MNRAS, 325, 726 
\bibitem[]{}Umeda H. \& Nomoto K. 2002, ApJ, 565, 385
\bibitem[]{}Whittet, D.C.B. 1992, {\it Dust in the Galactic Environment}, IOP Publishing, Great Britain 
\bibitem[]{}Wooden, D. H. 1997, in {\it The Astrophysical Implications of the Laboratory Study of Presolar Materials}, T. J. 
Bernatowicz and E. K. Zinner eds., The American Institute of Physics
\bibitem[]{}Woosley, S. E. \& Weaver, T. A. 1995, ApJSS, 101, 181
\end{thebibliography}
\end{document}